\documentclass[12pt]{iopart}

\usepackage{iopams}  
\usepackage{graphicx}
\usepackage{wrapfig}

\newcommand{\cN}{{\cal N}}
\def\r{{\boldsymbol r}}

\begin{document}

\title[Forward particle productions at RHIC and the LHC from CGC with local rcBK]
      {Forward particle productions at RHIC and the LHC
       from CGC within local rcBK evolution}

\author{H.~Fujii$^1$, K.~Itakura$^2$, Y.~Kitadono$^2$ and Y.~Nara$^3$}

\address{%
$^1$ Institute of Physics, University of Tokyo, Komaba, Tokyo 153-8902, Japan\\
$^2$ KEK Theory Center, IPNS, KEK, Tsukuba, Ibaraki 305-0801, Japan\\
$^3$ Akita International University, Yuwa, Akita City 010-1292, Japan}

\ead{hfujii@phys.c.u-tokyo.ac.jp}
\vspace{-2mm}

\begin{abstract}
In order to describe 
forward hadron productions in high-energy nuclear collisions,
we propose
a Monte-Carlo implementation of Dumitru-Hayashigaki-Jalilian-Marian 
formula with the unintegrated gluon distribution obtained numerically 
from the running-coupling BK equation.
We discuss influence of initial conditions for the BK equation
by comparing a model constrained by global fit of small-$x$ HERA data
and a newly proposed one from the ``running coupling MV model."
\end{abstract}

\vspace{-4mm}
\section{Phenomenology with rcBK equation}
Relativistic Heavy-Ion Collider (RHIC) at Brookhaven,
the QGP machine, has provided a new opportunity
to explore the physics of the dense gluonic state
at small Bjorken's $x$ in incoming hadrons, which is called 
Color Glass Condensate (CGC) 
\cite{Gelis:2010nm}.
Analyses of (e.g.) deeply inelastic scatterings (DIS) at HERA\cite{GBW,IIM}
and heavy-ion collisions at RHIC\cite{KLN}
support the phenomenology based on the CGC picture.

A recent highlight is the inclusion of the
running coupling effects in the 
Balitsky-Kovchegov equation (the rcBK equation)~\cite{Balitsky:2006wa,Kovchegov:2006vj},
which controls the $x$-evolution of the unintegrated gluon distribution 
$\widetilde \cN(k,y)$ in the incoming hadron ($k$ is 
the transverse momentum and $y=\ln (1/x)$ is the rapidity),
which
is obtained as the
Fourier-transform of the `dipole scattering amplitude' $\cN(r,y)$ 
with $r$ being the transverse size of a dipole. 
The rcBK equation has been confronted with the global analysis of
DIS\cite{AAMQS} as well as with the data obtained in proton-proton (pp),
proton-nucleus (pA)  and nucleus-nucleus~(AA) collisions
\cite{Albacete:2007sm,AlbaceteMarquet,AlbaceteDumitru,Albacete:2011fw}
(see \cite{AlbaceteQM11} for recent progress).
Our goal is to elaborate 
a detailed model for particle productions in forward regions 
at RHIC and the Large Hadron Collider (LHC),
by applying the gluon distribution obtained from the rcBK equation.

\section{Forward particle production}

\subsection{Initial conditions for rcBK equation}

With the DIS data for $x<x_0$ accumulated at HERA, one can constrain 
the numerical solution of the rcBK equation quite accurately.
This program was accomplished in \cite{AAMQS} by choosing 
an initial condition at $x=x_0=0.01$ as (e.g.)
\begin{eqnarray}
\cN(r,y_0)=1-\exp\left [
-\frac{(r^2 Q_{s0}^2)^{\gamma}}{4} \ln \left (
\frac{1}{\Lambda r} + e \right ) \right ]
\; .
\label{eq:IC}
\end{eqnarray}
Among several parameter sets now available,
we adopt here the AAMQS parameter set $h$ 
($Q_{s0}^2=0.1597$ GeV$^2$, $\gamma=1.118$, $C=2.47$,
 $\Lambda=0.241$ GeV, $\alpha_{\rm fr}$=1)~\cite{AAMQS}.

In addition, we will try a new initial condition, motivated by 
the McLerran-Venugopalan model with the running coupling 
modification (rcMV) \cite{Iancu:2004bx}:
\begin{equation}
\cN(r,y_0) = 
\frac{\alpha_{\mu}}{\alpha_s(r)}\left [
  1 - \exp\left (-\frac{r^2Q_{A}^2}{4}\right ) \right ] 
\; ,
\label{eq:ICrcMV}
\end{equation}
where $\alpha_s(r)=1/[b_0\ln(4C^2/(r^2\Lambda^2)+e)]$
with $b_0=(11-2 N_{\rm f}/3)/4\pi$ with $N_{\rm f}=3$,~and $\alpha_{\mu}$
is fixed by requiring that maximum value of $\cN(r,y_0)$ be unity.
We have set $\cN(r,y_0)=1$ for $r$ larger than this maximum point.
For the rcMV initial condition we use the parameters
$C=1$, $\Lambda=0.2$ GeV, $Q_{A}^2 \equiv Q_{s0}^2/\ln(Q_{s0}^2/\Lambda^2)$
with $Q_{s0}^2=0.2$ GeV$^2$ as a trial.

\subsection{DHJ formula}

In particle productions at very forward rapidities $y>0$ 
one can probe the saturation regime of one of the colliding hadrons.
In this situation a factorized formula for the cross section is proposed by
Dumitru, Hayashigaki, and Jalilian-Marian (DHJ)\cite{Dumitru:2005gt}:
\begin{eqnarray}
\frac{dN}{dy_h d^2 p_T} = \frac{K}{(2\pi)^2}
\sum_i \int _{x_F}^1 \frac{dz} {z^2}
\; {x_1 f_{i/p}(x_1,p^2_T)}
\; {\widetilde \cN \left(\frac{p_T}{z},x_2\right)}
\; {D_{h/i}(z, p_T^2)}
\; , 
\label{eq:DHJ}
\end{eqnarray}
where $f_{i/p}$ is the collinear distribution function for the 
large-$x_1$ parton $i$,  $\widetilde \cN$ describes the 
small-$x_2$ unintegrated gluon distribution,
and $D_{h/i}$ deals with the high-$p_T$ fragmentation
of a parton $i$ into a hadron $h$ with the momentum fraction $z$.
The parameter $K$ is introduced to absorb the higher-order contributions.
In AA collisions, $f_{i/p}$ and 
$\widetilde \cN$ are to be generalized to those for nuclei.

The formula (\ref{eq:DHJ}) was applied~\cite{Dumitru:2005gt}
to forward hadron productions in d-Au collisions at RHIC,
with adopting a certain CGC model for $\widetilde \cN$.
The larger gluon density in a nucleus is advantageous
for probing the saturation effects. 
Recently Albacete and Marquet\cite{AlbaceteMarquet} exploited the rcBK
equation to improve  the $\widetilde \cN$ part theoretically.
Assuming a homogeneous nucleus in the transverse plane,
they obtained the best description to date for the rapidity and
momentum dependence of the d-Au data at RHIC.

\subsection{Monte-Carlo implementation for nuclear collisions:
the MC-DHJ/rcBK model}

For more quantitative study, we need a nuclear model
including the impact-parameter dependence and initial fluctuations. 
In a MC implementation developed in \cite{DrescherNara}, 
nucleons are randomly sampled according to the Woods-Saxon density 
in an event-by-event basis, and the initial saturation momentum
$Q_{s0}^2$ in (\ref{eq:IC}) and (\ref{eq:ICrcMV})
is taken to be $Q^2_{s0}= N(\r_\perp)Q_{sp}^2$ 
at each transverse coordinate, where $N(\r_\perp)$ is the 
number of nucleons in the dense target within a transverse area $S_\perp$. 
We apply locally in the transverse plane 
the numerical solutions for the rcBK equation
with different initial values of $Q_{s0}^2$. 
Such an approach has been pursued within the so-called $k_\perp$ factorization
approximation for AA collisions, which reproduces successfully the centrality dependence
of the hadron multiplicity at RHIC and the LHC\cite{DrescherNara,AlbaceteDumitru,Albacete:2011fw}.

Here we combine the DHJ formula with the rcBK evolution in 
the MC implementation (MC-DHJ/rcBK). 
That is, we compute particle productions {\it at each transverse grid} 
$\r_\perp$
using the DHJ formula (\ref{eq:DHJ}) with $\widetilde \cN(k,y)$
numerically obtained from $Q_{s0}^2$ at grid $\r_\perp$
determined by the MC code:
\begin{eqnarray}
\frac{dN}{dy_h d^2p_T d \r_\perp}=T_d(\r_\perp) \times 
\left. \frac{dN}{dy_h d^2p_T }\right|_{{\rm DHJ}\; \r\perp}\, .
\end{eqnarray}
Here $T_d(\r_\perp)$ is the thickness function on the dilute side.
We stress as an advantage of this approach that 
{\it there is no more additional parameter} after fitting pp collisions. 
We comment also that the MC implementation allows us to study the 
initial fluctuations\cite{DrescherNara}.

\section{Results}

\begin{figure}
\includegraphics[width=0.48\textwidth]{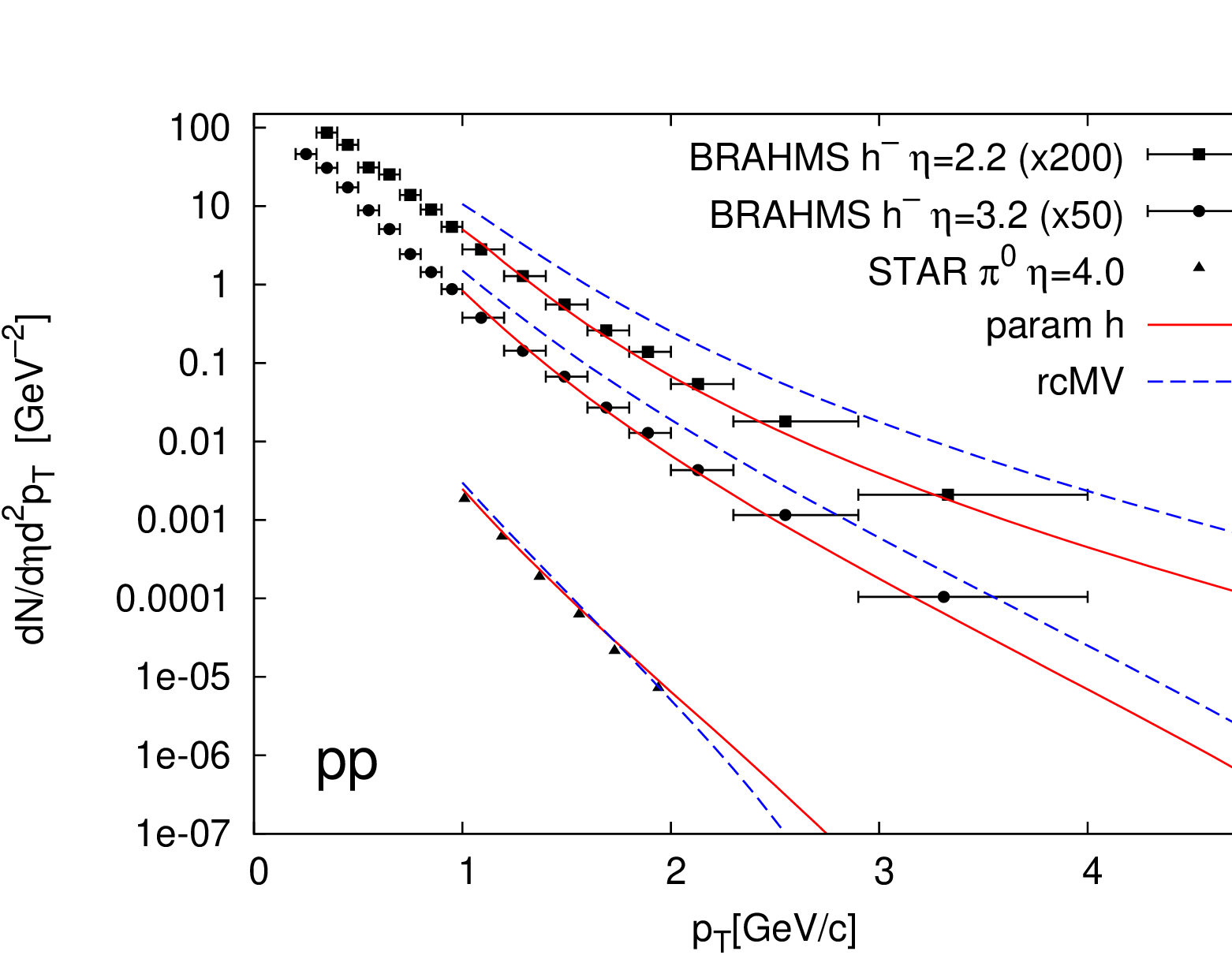}
\hfill
\includegraphics[width=0.48\textwidth]{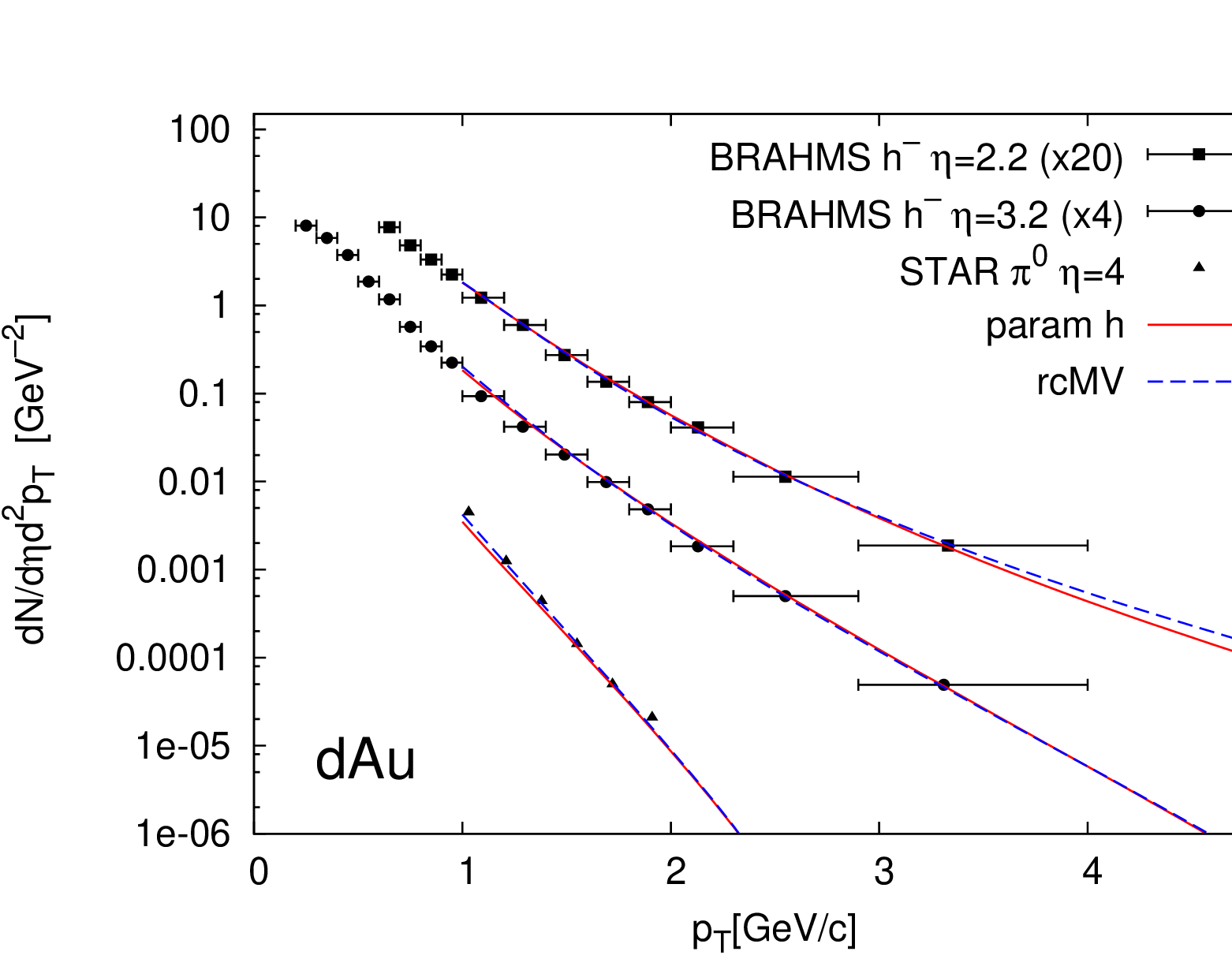}
\caption{Comparison of forward hadron spectra computed with 
the AAMQS set $h$ (solid) and rcMV (dashed)
to the data observed at RHIC.
The $K$ factor is chosen as $K=1.5$ for charged particles ($h^-$) and 
$K=0.5$ for $\pi^0$ in pp and d-Au collisions.}
\end{figure}

We use for $f_{i/p}$ and $D_{h/i}$
the CTEQ6M NLO PDF~\cite{CTEQ6} and 
DSS NLO fragmentation functions ~\cite{DSS}, respectively,
and set the factorization scale to $\mu^2 = p_T^2$.
We remark that 
an oscillation appears in $\widetilde N(k,y)$ 
for smaller $Q_{s0}^2$
when a sharp cutoff for the running coupling
$\alpha_s(r)=1/[b_0 \ln (4 C^2/r^2 \Lambda^2)]$
is adopted at $\alpha_{\rm fr}$. Thus we tried 
a smooth cutoff 
$\alpha_s(r)=1/[b_0 \ln (4 C^2/r^2 \Lambda^2)+a]$ where
constant $a$ is adjusted to make
$\alpha_s(r) \to 2$ as $r \to \infty$
in the rcBK evolution in the case of the rcMV initial condition.

In figure~1,  transverse momentum distributions of negatively charged hadrons $h^-$
at pseudo-rapidities $\eta=2.2$ and  3.2 from BRAHMS~\cite{BRAHMS}
and neutral pions $\pi^0$ at $\eta=4$ from STAR~\cite{exp:star} 
in pp and d-Au collisions at $\sqrt{s}=200$ GeV
are compared to our results.
The AAMQS set $h$ with $K=1.5$ (0.5)
describes the forward particle multiplicities 
of  $h^-$ ($\pi^0$) 
very nicely in pp and d-Au collisions at the same time without changing
any parameters in the model.
On the other hand, the rcMV 
initial condition with our current parameter set
with $K=1.5$ (0.5) for $h^-$ ($\pi^0$) 
leads to a good agreement with the data for d-Au collisions, 
%
\begin{wrapfigure}[16]{r}{0.5\textwidth}
\begin{center}
\vspace*{-\intextsep}
\includegraphics[width=0.5\textwidth]
{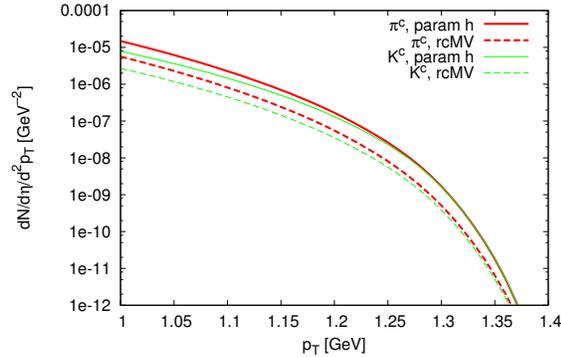}
\caption{Transverse momentum spectra 
of average of charged pions (bold) and of charged kaons (thin)
at $\eta=8.5$ 
for pp collisions at $\sqrt{s}=7$ TeV 
from MC-DHJ/rcBK model ($K$=1) with the AAMQS set $h$ (solid) and rcMV (dashed).
}
\end{center}
\end{wrapfigure}
but not for pp collisions.
It is very interesting to do global analysis of HERA data by
using rcMV initial condition and fix the parameters.
We will report more systematic analyses on the parameter dependence elsewhere.

Finally a result from
test calculations of the MC-DHJ/rcBK
at extremely forward rapidity $\eta=8.5$
for pp collisions at $\sqrt{s}=7$ TeV,
which is measured in LHCf experiment \cite{Adriani:2011nf},
is shown in figure~2.
One expects that dependence of the rcBK evolution on initial conditions
will become weaker at higher energies, but
we see that the result is still
relatively sensitive to the initial conditions.

\vspace{3mm}
\noindent
The authors would like to thank J.\ L.\ Albacete and A.\ Dumitru
for collaboration, and  
K.~Dusling, F.~Gelis, G.~Soyez and R.~Venugopalan for useful
conversations.

\end{document}